\documentclass[11pt]{article}
\usepackage{amsmath,amssymb}

\def\crampest{\medmuskip = 1mu plus 1mu minus 1mu}
\def\uncramp{\medmuskip = 4mu plus 2mu minus 4mu}
\def\ben{\begin{equation}}
\def\een{\end{equation}}

 \let\m=\mu \let\n=\nu  \let\p=\pi

\let\C=\Chi

\def\nn{\nonumber} \def\bd{\begin{document}} \def\ed{\end{document}}
\def\ds{\documentstyle} \let\fr=\frac \let\bl=\bigl \let\br=\bigr
\let\Br=\Bigr \let\Bl=\Bigl
\let\bm=\bibitem
\let\na=\nabla
\let\pa=\partial \let\ov=\overline
\newcommand{\be}{\begin{equation}}
\newcommand{\ee}{\end{equation}}
\def\ba{\begin{array}}
\def\ea{\end{array}}
\def\ft#1#2{{\textstyle{\frac{\scriptstyle #1}{\scriptstyle #2} } }}
\def\fft#1#2{{\frac{#1}{#2}}}
\def\del{\partial}
\def\vp{\varphi}
\def\sst#1{{\scriptscriptstyle #1}}
\def\oneone{\rlap 1\mkern4mu{\rm l}}
\def\td{\tilde}
\def\wtd{\widetilde}
\def\ie{{\it i.e.\ }}
\def\dalemb#1#2{{\vbox{\hrule height .#2pt
        \hbox{\vrule width.#2pt height#1pt \kern#1pt
                \vrule width.#2pt}
        \hrule height.#2pt}}}
\def\square{\mathord{\dalemb{6.8}{7}\hbox{\hskip1pt}}}
\newcommand{\ho}[1]{$\, ^{#1}$}
\newcommand{\hoch}[1]{$\, ^{#1}$}
\newcommand{\bea}{\setlength\arraycolsep{2pt} \begin{eqnarray}}
\newcommand{\eea}{\end{eqnarray}}
\newcommand{\ra}{\rightarrow}
\newcommand{\lra}{\longrightarrow}
\newcommand{\Lra}{\Leftrightarrow}
\newcommand{\bp}{\tilde \beta^\prime}
\newcommand{\tr}{{\rm tr} }
\newcommand{\Tr}{{\rm Tr} }
\def\0{{\sst{(0)}}}
\def\1{{\sst{(1)}}}
\def\2{{\sst{(2)}}}
\def\3{{\sst{(3)}}}
\def\4{{\sst{(4)}}}
\def\5{{\sst{(5)}}}
\def\6{{\sst{(6)}}}
\def\7{{\sst{(7)}}}
\def\8{{\sst{(8)}}}
\def\m{{\sst{(m)}}}
\def\n{{\sst{(n)}}}
\def\cA{{{\cal A}}}
\def\cB{{{\cal B}}}
\def\cF{{{\cal F}}}
\def\cG{{{\cal G}}}
\def\cH{{{\cal H}}}
\def\tV{\widetilde V}
\def\tW{\widetilde W}
\def\tH{\widetilde H}
\def\tE{\widetilde E}
\def\tF{\widetilde F}
\def\tA{\widetilde A}
\def\im{{{\rm i}}}
\def\tY{{{\wtd Y}}}
\def\ep{{\epsilon}}
\def\vep{{\varepsilon}}
\def\bD{{{\bar D}}}
\def\R{{{\mathbb R}}}
\def\C{{{\mathbb C}}}
\def\H{{{\mathbb H}}}
\def\CP{{{\mathbb C}{\mathbb P}}}
\def\RP{{{\mathbb R}{\mathbb P}}}
\def\Z{{{\mathbb Z}}}
\def\bA{{{\mathbb A}}}
\def\bB{{{\mathbb B}}}
\def\bC{{{\mathbb C}}}
\def\bD{{{\mathbb D}}}
\def\bE{{{\mathbb E}}}
\def\bZ{{{\mathbb Z}}}
\def\Re{{{\frak{Re}}}}
\def\Im{{{\frak{Im}}}}
\def\cosec{{\,\hbox{cosec}\,}}
\def\Gm{{\Gamma_{\!\! -}}}
\def\Gp{{\Gamma_{\!\! +}}}
\def\stan{{standard }}
\def\nonstan{{supernumerary }}
\def\p{{\partial}}
\def\kdel#1{{\fft{\del}{\del#1}}}

\def\bog{{Bogomolny }}
\def\om{{\omega}}

\newcommand{\nnr}{\nonumber \\}
\newcommand{\pd}{\partial}
\newcommand{\ud}{\textrm{d}}
\newcommand{\dTH}{T^{\prime \, 0}_\textrm{H}}
\newcommand{\dOi}{\Omega^{\prime \, 0}_i}

\newcommand{\tamphys}{\it George and Cynthia Woods Mitchell  Institute
for Fundamental Physics and Astronomy,\\
Texas A\&M University, College Station, TX 77843-4242, USA}

\newcommand{\auth}{
M. Cveti\v c\hoch{\$}, H. L\"u\hoch{\dagger\star}, 
and C.N. Pope\hoch{\dagger,\ddagger}
}

\begin{document}

\begin{flushright}
\hfill{
MIFP-09-35\ \ \ \ \ \ \ \  }\\
\hfill{
UPR-1206-T \ \ \ \ \ \ \  }\\
\end{flushright}

\begin{center}
{\large {\bf Non-Abelian Black Holes in $D=5$ Maximal Gauged
Supergravity}}

\vspace{15pt}
\auth

\vspace{10pt}

\hoch{\$}{\it Department of Physics and Astronomy,\\
University of Pennsylvania, Philadelphia, PA 19104, USA}

\vspace{10pt}
\hoch{\dagger}{\tamphys}

\vspace{10pt}


\hoch{\star}{\it Division of Applied Mathematics and Theoretical
Physics,\\
China Institute for Advanced Study,\\
Central University of Finance and Economics, Beijing, 100081, China
}

\vspace{10pt}

\hoch{\ddagger}{\it  DAMTP, Centre for Mathematical Sciences,
 Cambridge University,\\  Wilberforce Road, Cambridge CB3 OWA, UK}

\vspace{10pt}


\vspace{30pt}

\underline{ABSTRACT}
\end{center}

We investigate static non-abelian black hole solutions of anti-de
Sitter Einstein-Yang-Mills-Dilaton gravity, which is obtained as a
consistent truncation of five-dimensional maximal gauged
supergravity. If the dilaton is (consistently) set to zero, the
remaining equations of motion, with a spherically-symmetric ansatz,
may be derived from a superpotential. The associated first-order
equations admit an explicit solution supported by a non-abelian SU(2)
gauge potential, which has a logarithmically growing mass term.  In an
extremal limit the horizon geometry becomes AdS$_2\times S^3$.  If the
dilaton is also excited, the equations of motion cannot easily be
solved explicitly, but we obtain the asymptotic form of the more
general non-abelian black holes in this case.  An alternative
consistent truncation, in which the Yang-Mills fields are set to zero,
also admits a description in terms of a superpotential.  This allows
us to construct explicit wormhole solutions (neutral
spherically-symmetric domain walls).  These solutions may be
generalised to dimensions other than five.

\vspace{15pt}

\thispagestyle{empty}





\newpage

Explicit analytic solutions for non-abelian black hole or soliton
solutions of (gauged) supergravity theories are rare. The first such
example was that of the Chamseddine-Volkov BPS monopole \cite{CV} of
four-dimensional ${\cal N}=4$ gauged supergravity.  Its type IIB
embedding can be interpreted as D5-branes wrapped on $S^2$, and the
$D=4$, ${\cal N}=1$ dual field theory interpretation was given in
\cite{MN}.  This gravitating BPS soliton is supported both by the
$SU(2)$ gauge field and a scalar field.  The ground state of this
supergravity truncation does not have a constant scalar and so there
is no AdS$_4$ vacuum.  (For work on non-abelian BPS black holes and
Dirac-`t Hooft type monopoles in $D=4$, ${\cal N}=2$ ungauged
supergravity, see \cite{Ortin} and references therein.)

Numerical results, which provide evidence for the existence of regular
BPS monopole solutions in $D=5$, ${\cal N}=4$ gauged supergravity,
were presented in \cite{CVII}.  The lift to type IIB superstring
theory was interpreted as D5-branes wrapped on $S^3$, and the dual
$D=3$, ${\cal N}=1$ field theory interpretation was given in
\cite{MN}.  (These latter results can also be interpreted as those of
$D=7$, ${\cal N}=2$ gauged supergravity and its lifts to type IIB and
M-theory, analysed earlier in \cite{Gauntlett}. See also the review
\cite{Gauntlettrev} and references therein.)  Again these gravitating
BPS solitons are supported both by the $SU(2)$ gauge field and a
scalar field, and thus are not asymptotic to AdS$_5$.  (For further
numerical analysis of black hole and soliton solutions of $D=5$,
${\cal N}=4$ gauged supergravityi, see \cite{Radu} and references
therein. Non-abelian BPS solutions in five-dimensional ${\cal N}=2$ gauged
supergravity have been discussed recently in \cite{bell}.)
   
It is believed that maximal (${\cal N}=8$) gauged supergravity in
$D=5$ can be obtained from a Kaluza-Klein reduction of ten-dimensional
type IIB supergravity on $S^5$.  The only complete demonstrations so
far are for the consistency of the maximal abelian $U(1)^3$ truncation
\cite{tenauthor}, the ${\cal N}=4$ gauged $SU(2)\times U(1)$
truncation \cite{su2u1}, the scalar truncation in
\cite{scalar1,scalar2} and the $SO(6)$ truncation \cite{so6}.  The
form of the full metric reduction ansatz was conjectured in
\cite{metriconly}.

Non-abelian solutions in any of the five-dimensional gauged
supergravities that have a known embedding in type IIB supergravity
are of particular interest because they can be given a ten-dimensional
interpretation within string theory.  We can find an exact solution in
the $SU(2)\times U(1)$ gauged theory, whose type IIB embedding is
given in \cite{su2u1}, in which the $SU(2)$ Yang-Mills fields carry a
magnetic charge.  Unfortunately, however, the BPS condition implies
that the five-dimensional metric has the wrong signature.

In this paper we consider static non-abelian black hole solutions of
five-dimensional maximal (${\cal N}=8$) gauged supergravity.  We
present a consistent truncation of this theory whose bosonic sector
comprises gravity, $SU(2)\times SU(2)$ gauge fields, and a scalar
field whose potential has an AdS$_5$ minimum.  A further consistent
truncation to just an $SU(2)$ gauge symmetry, with a fixed
cosmological constant (related to the gauge coupling), results in
anti-de Sitter Einstein-Yang-Mills gravity.  We show that with the
assumption of a spherically-symmetric ansatz for the metric and
$SU(2)$ Yang-Mills potentials, the field equations for this truncated
system may be derived from a superpotential, and hence we can obtain
first-order equations of motion.  These give rise to an explicit
$SU(2)$ black hole solution which is asymptotic to $AdS_5$, but which
has a logarithimically divergent mass term as a consequence of the
non-vanishing (constant) $SU(2)$ gauge potential.  This solution was
in fact obtained previously in \cite{maeda}.\footnote{Other recent
  works on non-abelian solutions in various other Einstein-Yang-Mills
  systems can be found in \cite{nw1,nw2,nw3,nw4}.}  We find that it
admits an extremal limit, for which the $SU(2)$ gauge-potential
remains non-vanishing, and the horizon has the geometry AdS$_2\times
S^2$.  These results are intriguing, since the solutions have an
embedding into $D=5$ maximal gauged supregravity, and hence admit a
further lift to type IIB string theory.

Although one may expect that with a spherically-symmetric ansatz the
more general system with $SU(2)\times SU(2)$ gauge fields and a
dilatonic scalar should also admit a description in terms of a
superpotential, we have not succeeded in finding it in this case.  We
can, nevertheless, directly study the second-order equations of
motion, and investigate the asymptotic form for the more general
solutions with the additional ``scalar charge.''  We find evidence
that these non-abelian Yang-Mills solutions again describe black
holes, albeit again with logarithmically divergent mass.

We also find wormhole solutions of both $D=4$ and $D=5$ maximal gauged
supergravities. These are static neutral domain-wall solutions, which
are asympotic to AdS$_4$ and AdS$_5$ respectively.  However, in the
interior a scalar field diverges.  We obtain these solutions by
finding a superpotential, and then solving the assoicated first-order
equations of motion.  However, these solutions do not have
supersymmetric limits.

We start with the $SO(6)$ truncation of $D=5$, ${\cal N}=8$ gauged
supergravity.  It can be obtained from the $S^5$ reduction of the
$SL(2,\R)$-singlet sector of type IIB supergavity, for which the only
bosonic fields in ten dimensions are the metric and the self-dual
5-form.  The full non-linear ansatz was given in \cite{so6}.  The
five-dimensional theory consists of the metric, twenty scalars, which
are in the $20'$ respresentation of $SO(6)$ and are represented by the
symmetric unimodular tensor $T_{ij}$, with $i$ being a 6 of $SO(6)$,
together with 15 $SO(6)$ Yang-Mills gauge fields, represented by the
1-form potentials $A^{ij}$, antisymmetric in $i$ and $j$.  The
five-dimensional Lagrangian is given by \cite{so6}
\bea
{\cal L}_5 &=& R\, {*\oneone} - \ft14 T^{-1}_{ij}\, {*D T_{jk}}\wedge
T^{-1}_{k\ell}\, DT_{\ell i} - \ft14 T^{-1}_{ik}\,
T^{-1}_{j\ell}\, {* F^{ij}}\wedge F^{k\ell}
-V\, {*\oneone}\label{d5lag}\\
&& \!\!\! - \ft1{48}\, \ep_{i_1\cdots i_6}\,
\Big(F^{i_1 i_2}\, F^{i_3 i_4}\, A^{i_5 i_6} -
 g\, F^{i_1 i_2}\, A^{i_3 i_4}\, A^{i_5 j}\, A^{j i_6}
+\ft25 g^2\, A^{i_1 i_2} \, A^{i_3 j}\, A^{j i_4}\, A^{i_5
k}\, A^{k i_6} \Big)\,,\nn
\eea
where the potential $V$ is given by
\be
V = \ft12 g^2\, \Big(2 T_{ij}\, T_{ij} - (T_{ii})^2 \Big)\,.
\ee
The Yang-Mills field strength $F^{ij}$ and covariant derivative
$DT_{ij}$ are defined by
\be
F^{ij}=dA^{ij} + g A^{ik}\wedge A^{kj}\,,\qquad
DT_{ij}=dT_{ij} + g\, A^{ik} T_{kj} + g A^{jk} T_{ik}\,.
\ee

   We now perform a further truncation of $SO(6)$ to
$SU(2)\times SU(2)$, by setting
\bea
&&A^{12}=A^3\,,\quad A^{23}=A^1\,,\quad A^{31}=A^2\,,\qquad
A^{45}=\tilde A^3\,,\quad A^{56}=\tilde A^1\,,\quad A^{64}=\tilde A^2
\cr
&&T_{11}=T_{22}=T_{33}=X\,,\qquad T_{44}=T_{55}=T_{66}=X^{-1}\,,
\eea
with the remaining fields vanishing.  This truncation is consistent provided
that the additional constraint
\be
F^i\wedge \widetilde F^j=0\label{constraint}
\ee
is imposed, where
\be
F^i=dA^i + \ft12 g \epsilon^{ijk} A^j\wedge A^k\,,\qquad
\widetilde F^i=d\tilde A^i + \ft12 g \epsilon^{ijk} \tilde A^j\wedge
\tilde A^k\,.
\ee
The fields satisfy equations
of motion that can be derived from the Lagrangian
\bea
{\cal L} &=& R {*\oneone} - \ft32 X^{-2} {*dX}\wedge dX -
\ft12 X^{-2} {*F^i}\wedge F^i - \ft12 X^2 {*\widetilde F^i}\wedge
\widetilde F^i\nn\\
&&\qquad\qquad + \ft32 g^2 (X^2 + X^{-2} + 6)\,,
\eea
together with the constaint (\ref{constraint}).

       We may now look for spherically-symmetric static solutions,
by making the ansatz
\bea
ds_5^2 &=& -\alpha^2 dt^2 + d\rho^2 + \ft14 \beta^2
        (\sigma_1^2 + \sigma_2^2 + \sigma_3^2)\,,\cr
A^i &=&g^{-1}\gamma \sigma_i\,,\qquad \tilde 
A^i=g^{-1}\tilde \gamma \sigma_i
\,,
\eea
where the functions $\alpha, \beta, \gamma, \tilde\gamma$ and the
scalar $X$ are taken to depend only on the radial coordinate $\rho$.
The $\sigma_i$ are $SU(2)$ left-invariant 1-forms, satisfying
$d\sigma_i=-\ft12\epsilon^{ijk}\sigma_j\wedge \sigma_k$. Note that the
metric ansatz is invariant under $SO(4)\sim SU(2)_L\times SU(2)_R$
rotations of the $S^3$ level surfaces.  The Yang-Mills potentials, and
field strengths, are invariant under $SU(2)_L$, whilst they rotate
covariantly under $SU(2)_R$.  The energy-momentum tensor is therefore
invariant under the full $SO(4)$ action.

   The Yang-Mills equations imply that
\bea
(\alpha\beta \dot\gamma\, e^{-\phi})\dot{} -
\fft{4\alpha}{\beta} (\gamma\, \gamma-1)(2\gamma-1) e^{-\phi}
&=&0\,,\cr
(\alpha\beta \dot {\tilde\gamma}\, e^{\phi})\dot{} -
\fft{4\alpha}{\beta} \tilde\gamma\, (\tilde\gamma-1)(2\tilde\gamma-1)
e^{\phi} &=&0\,,\label{ymeom}
\eea
where we have defined
\be
X\equiv e^{\ft12 \phi}\,,
\ee
and a dot denotes a derivative with respect to $\rho$.
The scalar equation of motion is 
\bea
&&\fft{(\alpha\beta^3 \dot\phi)\dot{}}{2\alpha\beta^3} =
e^{-\phi}
\Big(g^2 - \fft{4\dot\gamma^2}{g^2\beta^2} -
\fft{16\gamma^2 (\gamma-1)^2}{g^2\beta^4}\Big) -
e^{\phi}\Big(g^2 - \fft{4\dot{\tilde\gamma}^2}{g^2\beta^2} -
\fft{16\tilde\gamma^2 (\tilde\gamma-1)^2}{g^2\beta^4}\Big)
\,,\label{scalareom}
\eea
and the Einstein equations are given by
\crampest
\bea
&&\fft{\ddot\alpha}{\alpha} + \fft{3\dot\alpha\dot\beta}{\alpha\beta}
= \Big(\fft{2\dot\gamma^2}{g^2\beta^2} +
\fft{8\gamma^2(\gamma-1)^2}{g^2\beta^4}\Big) e^{-\phi} +
\Big(\fft{2\dot{\tilde\gamma}^2}{g^2\beta^2} +
\fft{8\tilde\gamma^2(\tilde\gamma-1)^2}{g^2\beta^4}\Big) e^{\phi}
+g^2 (\cosh\phi +3)\,,\cr
&&\fft{\ddot\alpha}{\alpha} + \fft{3\ddot\beta}{\beta} =
-\Big(\fft{4\dot\gamma^2}{g^2\beta^2} -
\fft{8\gamma^2 (\gamma-1)^2}{g^2\beta^4}\Big)e^{-\phi} -
\Big(\fft{4\dot{\tilde \gamma}^2}{g^2\beta^2} -
\fft{8\tilde\gamma^2 (\tilde\gamma-1)^2}{g^2\beta^4}\Big)e^{\phi}
+g^2 (\cosh\phi+3) + \ft38 \dot\phi^2\,,\cr
&&\fft{2}{\beta^2} - \fft{\dot\alpha\dot\beta}{\alpha\beta} -
\fft{2\dot\beta^2}{\beta^2} - \fft{\ddot\beta}{\beta} =
\fft{8\gamma^2(\gamma-1)^2}{g^2\beta^4} e^{-\phi}
+\fft{8\tilde\gamma^2(\tilde\gamma-1)^2}{g^2\beta^4} e^{\phi}
-g^2 (\cosh\phi +3)\,,\label{einsteom}
\eea
\uncramp

   The constraint (\ref{constraint}) implies that
\be
\dot \gamma \tilde\gamma (\tilde \gamma-1) +
\dot {\tilde \gamma} \gamma (\gamma -1)=0\,.
\ee
 From this, it follows that
\be
\gamma \tilde \gamma = c (\gamma-1)(\tilde \gamma-1)\,,
\label{constraint1}
\ee
where $c$ is an integration constant.  Combining (\ref{constraint1}) and
(\ref{ymeom}), we obtain the first-order constraint
\be
\beta^2\, [c- (c-1) \gamma]\, \dot\phi\, \dot\gamma=
  (c-1)[4\gamma^2(\gamma-1)^2 - \beta^2\, \dot\gamma^2]
\,.\label{focon}
\ee
Finding the general solution to the equations of motion is likely to
be very difficult.  We can, however, obtain explicit exact solutions 
in some special cases.

       We first consider the case where the Yang-Mills fields are 
non-vanishing, and the integration constant 
$c$ in (\ref{constraint1}) is chosen to be $c=1$.  Equation 
(\ref{focon}) then implies that either $\dot\gamma =0$ or 
$\dot\phi=0$.  For $\dot\gamma=0$,
it follows from (\ref{constraint1}) that $\tilde\gamma=1-\gamma$, and
from (\ref{ymeom}) that the only non-trivial solution
is $\gamma=\ft12=\tilde \gamma$, in which case the scalar
equation (\ref{scalareom}) implies that we can also set $\phi=0$.
The reduced equations of motion can now be derived from an effective
Lagrangian $L=T - U$ with
\bea
T&=&\fft{6\alpha'\beta'}{\alpha\beta} +
\fft{6\beta'^2}{\beta^2}\,,\nn\\
U&=&-\ft{3}{32}\alpha^2 \beta^2 \Big(-\fft{1}{2g^2} +
\beta^2 + 2g^2 \beta^4 \Big)\,,
\eea
where a prime denotes a derivative with respect to $\eta$, defined
by $d\rho=\ft18 \alpha \beta^3 d\eta$.   

    Expressing the kinetic terms $T$ as
\be
T=\ft12 g_{ij} \fft{dX^i}{d\eta} \fft{dX^j}{d\eta}\,,\label{Texp}
\ee
where $X^i=(\alpha,\beta)$, we find that the potential $U$ can be
written in terms of a ``superpotential'' $W$ as
\be
U=-\ft12 g^{ij} \fft{\del W}{\del X^i} \fft{\del W}{\del X^j}\,,
\label{vwdef}
\ee
where
\be
W=\ft34\alpha\beta \sqrt{-M + \beta^2 + g^2 \beta^4 -
\fft{\log (g\beta)}{g^2}}\,.
\ee
The parameter $M$ is an integration constant, which can be chosen
arbitrarily.  The existence of the superpotential implies that the
second-order equations of motion are satisfied if the first-order
equations $dX^i/d\eta = g^{ij} \del W/(\del X^j)$ hold.  Thus we
obtain the equations
\bea
\dot\alpha &=&
\fft{\alpha (-1 +2 g^2 M  + 2 g^4 \beta^4 + 2 \log(g \beta))}{
2g^2\beta^2\sqrt{-M + \beta^2 + g^2 \beta^4 - g^{-2}\log(g\beta)}}\cr
\dot\beta &=& \beta^{-1}\sqrt{-M +  \beta^2 + g^2
\beta^4 - g^{-2} \log(g\beta)} \,.
\eea
These can be solved, giving
\bea
ds^2&=&-f dt^2 + \fft{dr^2}{f} + \ft14 r^2 (\sigma_1^2 +\sigma_2^2 +
\sigma_3^2)\,,\nn\\
A^i&=& \fft{1}{2g} \sigma_i = \tilde A^i\,,\qquad
f=1 + g^2 r^2 - \fft{M + g^{-2} \log(g r)}{r^2}\,.\label{ymbh}
\eea
Note that because the superpotential $W$ itself has the arbitrary
constant of integration $M$, the solution (\ref{ymbh}) of the
first-order equations is in fact the most general solution also of the
original second-order field equations.  The solution describes an
$SU(2)$ Yang-Mills black hole with logarithmically divergent mass.
The horizon is located at the largest root of the function $f$.

   The solution has an extremal limit, for which $f(r)$ and $f'(r)$
vanish simultaneously at $r=r_0$, when
\be
M=\fft{1 + \sqrt5 + \log(8(7 + 3\sqrt5))}{8g^2}\,,
\ee
and
\be
g^2r_0^2=\ft14 (\sqrt5-1)\,.
\ee
The near-horizon geometry is then a direct product of AdS$_2\times S^3$,
with the metric given by
\bea
ds^2 &=& \alpha^2 ds_2^2 + \beta^2 d\Omega_3^2\,,\cr
\alpha^2&=&\fft{5-\sqrt5}{40g^2}\,,\qquad
\beta^2=\fft{\sqrt5-1}{4g^2}\,.
\eea
It is worth pointing out that it is the Yang-Mills fields that are
responsible for the occurrence of the logarithmic term in the metric
(\ref{ymbh}). If the Yang-Mills fields are instead set to zero, the
logarithmic term disappears and the solution becomes the
AdS-Schwarzschild black hole.  That non-supersymmetric solutions such
as the Schwarzschild black hole can be obtained from first-order
equations derived from a superpotential was previously observed in
\cite{lupova}.

   The solution can be lifted back to $D=10$, by using the 
reduction ansatz given in \cite{so6}. The ten-dimensional
metric is given by
\bea
d\hat s_{10}^2 &=& ds_5^2 + \fft{1}{g^2} \Big\{ [d\mu_1 + 
 \ft12 (\mu_2\sigma_3 - \mu_3\sigma_2)]^2
+ [d\mu_2 + \ft12 (\mu_3\sigma_1 - \mu_1\sigma_3)]^2\nn\\
&& + 
[d\mu_3 + \ft12 (\mu_1\sigma_2 - \mu_2\sigma_1)]^2
+ [d\mu_4 + \ft12 (\mu_5\sigma_3 - \mu_6\sigma_2)]^2\nn\\
&&+ [d\mu_5 + \ft12 (\mu_6\sigma_1 - \mu_4\sigma_3)]^2 +
[d\mu_6 + \ft12 (\mu_4\sigma_2 - \mu_5\sigma_1)]^2\Big\}\,,
\eea
where $\mu_i \mu_i=1$.  The RR 5-form field takes a somewhat involved
structure, supported by the five-sphere coordinates; it can be readily
derived from the expressions given in \cite{so6}.  Thus the Type IIB
solution is supported by the RR 5-form with a specific non-abelian
deformation of the five-sphere coordinates, resulting in an
asymptotically AdS$_5$ space-time supported by D3-brane fluxes.  This
structure is different from those of \cite{Maldanunez,Gauntlettrev}
where a five-brane wraps a 2-cycle in the internal Ricci-flat space,
resulting in a $D=4$ ${\cal N}=1$ dual field theory on the
world-volume of the five-brane.

We may also consider the more general case of non-abelian solutions
where the dilatonic scalar is also excited.  For simplicity, we
consider only the case for $\gamma =\tilde \gamma=\ft12$, so that the
Yang-Mills equations are trivially satisfied.  The equations of motion
for the spherically-symmetric ansatz can then be derived from the
Lagrangian $L=T-U$, with
\bea
T&=& \fft{6\alpha'\beta'}{\alpha\beta} +
\fft{6\beta'^2}{\beta^2}-\ft38 {\phi'}^2\,,\nn\\
U&=& -\ft{3}{32}\alpha^2\beta^2\Big(-\fft1{2g^2} \cosh\phi +
  \ft12 g^2 \beta^4 \cosh\phi +\beta^2 +\ft32 g^2 \beta^4\Big)\,.
\eea
Reading off $g_{ij}$ using (\ref{Texp}), and then expressing $U$ in 
terms of a superpotential $W$ as in (\ref{vwdef}), we find that $W$
is determined by the equation 
\be
\beta \fft{\del \hat W^2}{\del\beta} -
  8 \Big(\fft{\del\hat W}{\del\phi}\Big)^2 =
  3 g^2 \beta^4 + 2 \beta^2  +\Big(g^2\beta^4 -\fft{1}{g^2}\Big)\cosh\phi\,,
\ee
where
\be
W(\alpha,\beta,\phi)= \fft{3\alpha\beta}{4}\, \hat W(\beta,\phi)\,.
\ee
We have not been able to solve this equation explicitly.

Although we are unable to obtain the exact solution, we may
nevertheless consider its large $r$ expansion, which we find to be
given by
\bea
&&ds^2 = -N^2 dt^2 + \fft{dr^2}{f} + \ft14 r^2 (
\sigma_1^2 + \sigma_2^2 + \sigma_3^2)\,,\qquad
A^i = \fft{1}{2g} \sigma_i = \tilde A^i\,,\cr
&&N^2=1 + g^2 r^2 - \fft{\xi-1}{g^2 r^2} +
\fft{q^2(8\xi-3)}{32 g^2 r^6} - \fft{q^2(40\xi-1)}{800 g^4 r^8}\cr
&& \qquad\qquad+\fft{q^2\Big(1200 \xi^2 - 60 (13 + 20 g^4 q^2) \xi
- 13 + 550 g^4 q^2\Big)}{9600 g^6 r^{10}}\cr
&&\qquad\qquad-\fft{q^2 \Big(4492320 \xi^2- 56 (56713 + 25200 g^4  q^2)
\xi + 3 (63461 + 64400 g^4  q^2 )\Big) }{39513600 g^8  r^{12}}+
\cdots\,,\cr
&&\fft{f}{N^2}=1 + \fft{q^2}{r^4} + \fft{q^2(24\xi-1)}{32g^4 r^8} -
    \fft{q^2(120\xi-13)}{225g^6 r^{10}}\cr
&&\qquad\qquad + \fft{q^2 \Big(2016 \xi^2 - 48 (3 + 8 g^4 q^2) 
\xi -1 + 160 g^4 q^2\Big)}{3072 g^8 r^{12}} + \cdots\,,\cr
&&\sinh\ft12\phi=\fft{q}{r^2}\Big[1 + \fft{\xi}{4g^4 r^4} -
\fft{12 \xi -1}{72 g^6 r^6} + \fft{288 \xi^2 - 128 g^4 q^2 \xi -1 +
32 g^4 q^2}{2048 g^8}\cr
&&\qquad\qquad\fft{631200 \xi^2 - 40 (8053 + 3600 g^4  q^2 ) \xi +
9 (4303 + 4400 g^4  q^2 )}{2880000g^{10}r^{10}}\cr
&&\qquad\qquad\fft{1}{82944000 g^{12}r^{12}}\Big(8100000 \xi^3 +
28800 (256 - 285 g^4  q^2 ) \xi^2\cr
&&\qquad\qquad\qquad + 15
(-508631 + 141600 g^4  q^2  + 172800 g^8  q^4 ) \xi\cr
&&\qquad\qquad\qquad\qquad
1093091 + 18000 g^4  q^2  - 756000 g^8  q^4\Big)
 +\cdots\Big]\,,
\eea
where $\xi$ is given by
\be
\xi=1 + g^2 M  +\log(gr)\,.
\ee
Although we have not obtained the exact solution, the form of the
large-$r$ expansion indicates that it describes a black hole, at least
provided that $q$ is sufficiently small or that $M$ is sufficiently
large.  To see this, we note that for the case $q=0$, the solution
reduces to the exact one that we discussed earlier.  The horizon is
located at $r_+$, where $M=r_+^2 + r_+^4- \log r_+$.  (Here we set
$g=1$ for simplicity.)  With $q\ne 0$, the horizon is shifted to
$r_+'$ where $r'_+$ is defined by $N(r_+')=0$.  It is straightforward
to see that
\be
r'_+-r_+ = {\cal O}\Big(\fft{q^2}{r_+^3}\Big)\,.
\ee

We may alternatively consider a special case where the Yang-Mills
fields are set to zero, so that the solution then involves only the
metric and a scalar field.  The scalar potential fits into the general
pattern discussed in the appendix.  Following the discussion in the
appendix, we find that there exists a domain wall solution
\bea
ds^2_5&=&-(k+g^2 r^2) dt^2 + \fft{dr^2}{(k + g^2 r^2)
(1 + \fft{q^2}{r^4})} + r^2 d\Omega_{3,k}^2
\,,\cr \sinh\ft12\phi &=& \fft{q}{r^2}\,,\label{d5dw1}
\eea
where $k=1$, 0 or $=-1$ for spherical, flat or hyperbolic spatial
sections.  This solution can also be lifted back to $D=10$, giving
rise to a solution of type IIB supergravity, using the results of
\cite{so6}:
\bea
d\hat s_{10}^2&=& \Big(X c^2 + \fft{s^2}{X}\Big)^2 \Big\{ ds_5^2 +
g^{-2} \Big(d\theta^2 + \fft{c^2}{X^2 c^2 + s^2} d\Omega_2^2 +
\fft{s^2}{s^2 X^{-2} + c^2} d\widetilde \Omega_2^2\Big)\Big\}\,,\cr
\hat G_\5 &=& g (X^2 c^2 + X^{-2} s^2 +3) \,\ep_\5 -\fft{2sc}{g}\,
X^{-1}\, {*dX}\wedge
d\theta\,,\nn\\
X&=& \fft{q}{r^2} + \sqrt{1 + \fft{q^2}{r^4}}\,,\qquad
c=\cos\theta\,,\qquad s=\sin\theta\,,
\eea
where the self-dual 5-form is given by $\hat F_\5=\hat G_\5 + 
{\hat *\hat G}_\5$.

The domain wall solution (\ref{d5dw1}) is massless and has a naked
singularity at $r=0$.  As we have shown in the appendix for general
class of such a domain walls in arbitrary dimensions, we can also add
a mass term.  The exact form of the solution is unknown, but the large
$r$ expansion can be obtained straightforwardly.  Here we present the
solution in higher orders:
\bea
N^2&=& k + g^2 r^2  - \fft{M}{r^2} +
\fft{M q^2}{4r^6} - \fft{k M q^2}{20 g^2 r^8} +
\fft{M q^2(4 k^2 + 15 g^2 M - 15 g^4 q^2)}{120 g^4 r^{10}}
+\cdots\,,\cr
\fft{f}{N^2} &=& 1 + \fft{q^2}{r^4} + \fft{3M q^2}{4g^2 r^8} -
\fft{8k M q^2}{15 g^4 r^{10}} + \cdots\,,\cr
\sinh\ft12\phi&=&1 + \fft{M}{4g^2 r^4} - \fft{k M}{6g^4 r^6}
+\fft{M(8k^2 + 9 g^2 M - 4 g^4 q^2)}{64g^6 r^8}\nn\\
&& -\fft{k M (120 + 263 g^2 M - 60 g^4 q^2)}{1200g^8r^{10}}
+ \cdots\,.
\eea
As in our previous discussion of the non-abelian black holes, since
the $q=0$ solution here describes the Schwarzschild-AdS black hole,
and the effect of the $q$ parameter is to modify terms at higher
orders in $1/r$, it follows that the solution with $q\ne0$ will still
describe a black hole, at least if $q$ is sufficiently small.

In summary, we have in this paper studied the system of equations that
arises from a consistent trunction of five-dimensional maximal gauged
supergravity, in which just the metric, a dilatonic scalar, and the
gauge fields of $SU(2)\times SU(2)$ are retained.  Consistency
requires that the gauge fields satisfy the constraint
(\ref{constraint}).  Our focus has been on seeking
spherically-symmetric solutions to this system.  In the two special
cases where either the dilaton is set to zero and a further truncation
to $SU(2)$ gauge symmetry is performed, or where the dilaton is
retained but the gauge fields are set to zero, we have been able to
describe the system in terms of a superpotential.  This allows us to
obtain first-order equations of motion, which can be straightforwardly
solved analytically.  The first of these cases leads to a non-abelian
black-hole solution, with logarithmically-diverging mass, which was
first obtained in \cite{maeda}.  The second case gives rise to a
spherically-symmetric domain wall solution.  In the more general
situation where both the Yang-Mills fields and the dilaton are
excited, we have not succeeded in describing the system in terms of a
superpotential.  Nevertheless, we have studied the asymptotic
behaviour of spherically-symmetric solutions, and found that more
general non-abelian black holes arise here also.  This discussion is
extended, in an appendix, to gravity plus dilaton systems in arbitrary
dimensions.

\bigskip

\noindent{\bf Acknowledgments}

M.C., H.L. and C.N,P are grateful to Sheridan Lorenz and the Mitchell
Family Foundation for hospitality at Cook's Branch Conservancy, and
H.L. and C.N.P. are grateful to the Department of Physics and
Astromony at the University of Pennsylvania for hospitality, during
the course of this work.  The work of M.C. is supported by DOE grant
DE-FG05-95ER40893-A020, NSF RTG grant DMS-0636606 and the Fay R. and
Eugene L. Langberg Chair.  The work of C.N.P. is supported in part by
DOE grant DE-FG03-95ER40917.

\appendix
\section{A spherically symmetric domain wall solution}

    In this appendix, we consider a general class of
$D$-dimensional Lagrangians, given by
\bea
{\cal L}_D &=& e(R - \ft12(\del\phi)^2 - V)\,,\cr
V&=&-(D-2) g^2 \Big[ D-2 + \cosh\Big(\sqrt{\fft{2(D-3)}{D-2}}
\, \phi\Big) \Big]\,.
\eea
The scalar potential can be expressed in terms of a superpotential $w$
as
\be
V=\Big(\fft{dw}{d\phi}\Big)^2 - \fft{D-1}{2(D-2)} w^2\,,\qquad
w=\sqrt2\,(D-2) g \cosh\Big(\sqrt{\fft{D-3}{2(D-2)}}\,\phi\Big)\,.
\ee
Consider the spherically-symmetric ansatz
\be
ds^2 = - \alpha^2 dt^2 + d\rho^2 + \beta^2 d\Omega_{n,k}^2\,,
\ee
where $n=D-2$, and $k=-1, 1$ and 0, corresponding to hyperbolic, flat
and sphere.   The scalar and Einstein equations of motion are
given by
\bea
\fft{(\alpha\beta^n \dot \phi)\dot{}}{\alpha \beta^n}
- \fft{dV}{d\phi} &=&0\,,\cr
\fft{\ddot\alpha}{\alpha} + \fft{n\,\dot\alpha\dot\beta}{\alpha\beta}+
\fft{V}{n}&=&0\,,\cr
-\fft{\ddot\alpha}{\alpha} - \fft{n\,\ddot\beta}{\beta} -
\fft12\dot\phi^2-\fft{V}{n}&=&0\,,\cr
\fft{(n-1)k}{\beta^2} -\fft{\dot\alpha\dot\beta}{\alpha\beta} -
\fft{(n-1)\dot\beta^2}{\beta^2} - \fft{\ddot\beta}{\beta}
-\fft{V}{n} &=& 0\,.
\eea
These can be derived from the Lagrangian ${\cal L}=T-U$, where
the kinetic and potential terms are given by
\bea
T&=&\fft{2(D-2)\alpha'\beta'}{\alpha\beta} +
\fft{(D-2)(D-3)\beta'^2}{\beta^2} -\ft12\phi'^2\,,\cr
U&=&\alpha^2\beta^{2(D-3)} (\beta^2 V - (D-2)(D-3)k)\,.
\label{scalartu}
\eea
Here, a prime denotes a derivative with respect to a new radial
coordinate $\eta$, which is defined by $d\eta = \alpha \beta^n d\rho$.
We find that with $U$ given by (\ref{scalartu}), there exists a
superpotential $W$, \`a la (\ref{vwdef}), given by
\be
W=2(D-2)\alpha\beta^{D-3} \sqrt{k + g^2\beta^2}
\cosh\Big(\sqrt{\fft{D-3}{2(D-2)}}\,\phi\Big)\,.\label{suppot}
\ee

   The resulting first-order equations of motion are
\bea
\dot\alpha &=& \fft{g^2\alpha\beta \cosh\Big(\sqrt{
\fft{D-3}{2(D-2)}}\,\phi\Big)}{\sqrt{k+g^2\beta^2}}\,,\qquad
\dot\beta = \sqrt{k+g^2\beta^2}\,\cosh\Big(\sqrt{
\fft{D-3}{2(D-2)}}\, \phi\Big)\,,\cr
\dot\phi &=& -\fft{\sqrt{2(D-3)(D-2)}}{\beta}
\sqrt{k + g^2 \beta^2}\, \sinh\Big(\sqrt{\fft{D-3}{2(D-2)}}\,
\phi\Big)\,.\label{genscalarfo}
\eea
Their solution gives
\bea
ds_D^2 &=& - (k + g^2 r^2) dt^2 + \fft{dr^2}{
(k + g^2 r^2)\Big(1 +
\fft{q^2}{r^{2(D-3)}}\Big)} + r^2 d\Omega_{D-2, \epsilon}^2\,,\cr
&&\sinh\Big(\sqrt{\fft{D-3}{2(D-2)}}\,\phi\Big)=
\fft{q}{r^{D-3}}\,.
\eea

The supersymmetry transformation rules for such a system
are given by
\be
\delta \psi_M = D_M \varepsilon - \fft{1}{2\sqrt2\, (D-2)} w\,
\varepsilon\,,\qquad
\delta \lambda = \fft{1}{2\sqrt2} \del_{M} \phi\, \Gamma^M \varepsilon
+ \fft12 \fft{d w}{d\phi}\, \varepsilon\,.\label{genscalarso}
\ee
It follows from $\delta\lambda=0$ that the existence of supersymmetry
would imply that
\be
\dot\phi = \pm \sqrt{2(D-3)(D-2)}\, g \sinh\Big(\sqrt{
\fft{D-3}{2(D-2)}}\, \phi\Big)\,.
\ee
Comparing this to the last equation in (\ref{genscalarfo}), the only
solution with supersymmetry is when $k=0$.  In fact it is easy to see
that (\ref{genscalarso}) cannot be supersymmetric for $k\ne0$.  If it
were, there would be a smooth limit when $g=0$, and this would lead to
a solution supported by the metric and a free scalar only, which could
not possibly be supersymmetric.

The scalar potential for $D=4$ occurs in four-dimensional ${\cal
  N}=4$, $SO(4)$ gauged supergravity.  The explicit reduction ansatz
that gives this theory from $D=11$ supergravity was found
in\cite{4from11}.  We can use the ansatz to lift the ($k=1$) solution
back to $D=11$, giving
\bea
d\hat s_{11}^2 &=& \Delta^{\ft13} \Big\{ -(1+g^2 r^2) dt^2 +
\fft{dr^2}{(1 + g^2 r^2)(1 + \fft{q^2}{r^2})} + r^2 d\Omega_2^2\cr
&&+ \fft{4}{g^2} \Big( d\xi^2 + \fft{c^2}{c^2X^2 + s^2} d\Omega_3^2 +
\fft{s^2}{s^2 X^{-2} + c^2} d\widetilde \Omega_3^2\Big)\Big\}\,,\cr
\hat F_\4 &=& -g (2 + X^2 c^2 + X^{-2} s^2) \epsilon_\4 - \fft{4sc}{g}
X^{-1} {*dX}\wedge d\xi\,,\cr
\Delta &=& (c^2 X^2 + s^2) (s^2 X^{-2} + c^2)\,,\qquad
X=\fft{q}{r} + \sqrt{1 + \fft{q^2}{r^2}}\,,
\eea

The domain wall solutions we obtained so far have zero mass, and a
naked singularity.  It is possible to add mass term, such that the
second-order equations of motion (but no longer the first-order
equations (\ref{genscalarfo}) following from the superpotential
(\ref{suppot})) are still satisfied.  The solution then develops a
horizon.  We are unable to find the exact solution for this case.
However, the large $r$ expansion of the solution can be obtained, and
is given by
\bea
ds_D^2 &=& -N^2 dt^2 + \fft{dr^2}{f} + r^2 d\Omega_{D-2, k}^2\,,\cr
&&N^2 = k + g^2 r^2 - \fft{M}{r^{D-3}} +
\fft{(D-1)M q^2}{2(3D-7)r^{3(D-3)}} + \cdots\,,\cr
&&\fft{f}{N^2} = 1 + \fft{q^2}{r^{2(D-3)}} +
\fft{4(D-2)(D-3) M q^2}{(D-1)(3D-7) r^{3D-7}} + \cdots\cr
&&\sinh\Big(\sqrt{\fft{D-3}{2(D-2)}}\,\phi\Big)
= \fft{q}{r^{D-3}} \Big(1 + \fft{(D-3)M}{2(D-1) g^2 r^{D-1}} + \cdots\Big)\,.
\eea
For vanishing $q$, the solution becomes the Schwarzschild AdS black
hole, whilst for vanishing $M$, it reduces to the singular domain wall
described earlier.  Since the effect of introducing $q$ is to modify
the behaviour only at higher inverse powers of $r$, it is clear that
for large enough $M$ or small enough non-zero $q$, the solution still
describes a black hole.

\end{document}